\documentclass[12pt]{article}
 \def\s{\sigma}
 \def\d{\delta}

\begin{document}

 \title{New critical frontiers for the Potts and percolation models}
 \vskip 1cm
\author{
F. Y. Wu \\ Department of Physics \\
Northeastern University, Boston, Massachusetts 02115, U.S.A.}

\date{}
\maketitle

\abstract{We obtain the critical threshold for a host of Potts and
percolation models on lattices having a structure which permits a
duality consideration. The  consideration generalizes the recently
obtained thresholds of Scullard and Ziff for bond and site
percolation on the martini and related lattices to the Potts model
and to other lattices.}

 \vskip 1cm


 \newpage

\section{Introduction}
The Potts model \cite{potts} has been in the forefront of active
research for many years. Despite  concerted efforts, however,
very few exact results are known \cite{wu}. Unlike the Ising
model for which the exact solution is known for all
two-dimensional lattices, the relatively simple question of
locating the critical frontier of the Potts model has been resolved
only for the square, triangular, and honeycomb lattices
\cite{bta,hkw,wu1}.
 The determination of the Potts critical
frontier for other two-dimensional lattices
 has remained very much an open problem \cite{wuhu}.

\medskip
In two recent papers using a star-triangle relation and a dual
transformation, Scullard \cite{scullard} and Ziff \cite{ziff} succeeded
to determine the critical thresholds of site and bond percolation
processes for several new two-dimensional lattices. As percolation
problems are realized in the $q=1$ limit of the $q$-state Potts
model \cite{kf,kw}, the new percolation results suggest the
possibility that  similar thresholds can also be determined for the Potts model.
In this paper we report this extension. We derive more generally the exact
critical frontier of the Potts model for a large class of
two-dimensional lattices including those considered in
\cite{scullard,ziff}, and obtain the corresponding percolation
thresholds.

\medskip
Consider a lattice having the structure shown in Fig. 1, where
each shaded triangle denotes a network connected to its exterior
through 3 spins $\s_1,\s_2,\s_3$. It was established by Baxter,
Temperley and Ashley \cite{bta} using an algebraic approach that
this Potts model possesses a duality relation and a self-dual
trajectory. A graphical proof of the duality relation was later
given by Wu and Lin \cite{wulin}, and subsequently Wu and Zia
\cite{wuzia} established rigorously that in the ferromagnetic
regime of the parameter space the critical threshold is indeed the
self-dual trajectory.

 \medskip
Specifically, write the Boltzmann factor for the shaded triangle
as
\begin{equation}
F(\s_1,\s_2,\s_3)=A+B_1\d_{23}+B_2\d_{31}+B_3\d_{12}+C\,\d_{123}\label{F}
\end{equation}
 where
$\d_{ij}=\d_{\s_i,\s_j},\d_{ijk}=\d_{ij}\d_{jk}\d_{ki}$. Then
the model possesses a duality relation in the parameter space
$\{A, B_1,B_2,B_3,C\}$.  In the ferromagnetic regime
\begin{equation}
B_1+B_2+B_3+C>0,\hskip1cm B_i+B_j+C>0,\quad i\not=j\label{ferro}
\end{equation}
the critical frontier of the Potts model is given
by the self-dual trajectory
\begin{equation}
qA-C=0.\label{cf}
\end{equation}
By realizing the shaded network as a simple triangle, for example,
one recovers from (\ref{cf}) the critical point for the Potts and
bond percolation models on the square, triangular, and honeycomb
lattices \cite{bta}.  Another realization of the Boltzmann factor
(\ref{F}) is the random cluster model \cite{kf} with 2- and 3-site
interactions \cite{wulin}. The isotropic version of the random cluster
 model has been analyzed very recently by Chayes and Lei \cite{chayeslei} who
established on a rigorous ground the duality relation and the
self-dual trajectory (\ref{cf}). Our new results concern with
other realizations of (\ref{F}).

\section{The martini lattice}

Consider the network shown in Fig. 2 as an instance of the shaded
triangle in Fig. 1. This gives rise to the martini lattice shown
in Fig. 3 \cite{scullard,ziff}. The Boltzmann factor for the
network is
\begin{eqnarray}
F(\s_1,\s_2,\s_3) &=&\sum_{\{\s_4,\s_5,\s_6\}=1}^q{\rm exp}
\Big[V_1\d_{14}+V_2\d_{25}+V_3\d_{36}\nonumber \\
&&\quad \quad +W_1\d_{56}+W_2\d_{46}+W_3\d_{45}+M\d_{456}\Big],
\label{F1}
\end{eqnarray}
where $V_i$ nd $W_i$ are 2-site Potts interactions and $M$ a
3-site interaction.

\medskip
It is straightforward to cast (\ref{F1}) in the form of (\ref{F})
\cite{mrw,kingwu} to obtain
\begin{eqnarray}
A &=&
v_1v_2v_3+v_1v_2
(q+w_1+w_2)+v_2v_3(q+w_2+w_3)+v_3v_1(q+w_3+w_1)
\nonumber \\
&& \quad +(q+v_1+v_2+v_3)\Big[q^2+q(w_1+w_2+w_3) +h\Big]\nonumber\\
B_i &=& v_jv_k\Big[h+(q+v_i)w_i\Big] , \quad i\not=j\not=k\not=j
\nonumber \\
C&=& v_1v_2v_3h , \label{abc}
\end{eqnarray}
where
\begin{eqnarray}
v_i &=&e^{V_i}-1, \hskip 2cm
w_i = e^{W_1}-1 \nonumber \\
h&=& e^{M+W_1+W_2+W_3} -e^{W_1}-e^{W_2}-e^{W_3}+2.
\end{eqnarray}
As alluded to in the above, in the ferromagnetic regime $W_i\geq
0, V_i\geq 0, M\geq 0$ satisfying (\ref{ferro}),
the critical frontier of this Potts model
is  the self-dual trajectory (\ref{cf}) which now reads
\begin{eqnarray}
&& q(q+v_1+v_2+v_3)\Big[q^2+q(w_1+w_2+w_3)+h\Big] \nonumber \\
&& \hskip 1cm +q\Big[v_1v_2v_3+v_1v_2(w_1+w_2+q)+v_2v_3(w_2+w_3+q)
 \nonumber\\
&& \hskip 1cm +v_3v_1(w_3+w_1+q)\Big]-v_1v_2v_3h=0. \label{3spin}
\end{eqnarray}
The critical frontier (\ref{3spin})
is a new result for the Potts model.

\medskip
For $M=\infty$ one retains only terms linear in $h$ and
(\ref{3spin}) reduces to the critical frontier
$q^2+q(v_1+v_2+v_3)=v_1v_2v_3$ of the honeycomb lattice.
For $M=0$, $V_1=V_2=V_3=V,\ W_1=W_2=W_3=W$, which is the isotropic model with
pure 2-site interactions,
 (\ref{3spin}) becomes
\begin{equation}
q(q+3v)(q^2+3qw+3w^2+w^3)+qv^2(v +
6w+3q)-v^3(3w^2+w^3)=0,\label{mPotts}
\end{equation}
where $v=e^V-1,\, w=e^W-1$.
 For $w=v$ it reduces further to the expression
\begin{equation}
q^4+6q^2v+q^2v^2(15+v)+qv^3(16+3v)-v^5(3+v)=0.\label{martinicritical}
\end{equation}

\medskip
One variation of the martini lattice
is the A lattice \cite{scullard,ziff} shown in Fig. 4(a)
obtained from the martini lattice by setting $v_1 =
\infty, v_2=v_3=v, w_1=w_2=w_3 =w$. This
gives rise to the Potts critical frontier
\begin{eqnarray}
&&q^3 +q^2(2v+3w) + q(v^2+4wv +3w^2+w^3) -v^2(3w^2+w^3)=0,\nonumber \\
&&\hskip 6cm \quad A {\rm \>\>
lattice}, \label{mAPotts}
\end{eqnarray}
 Another variation of the martini
lattice is the B lattice \cite{scullard,ziff} shown in Fig.
4(b) obtained  from the martini lattice by setting
$v_2=v_3=\infty, v_1=v, w_1=w_2=w_3=w$. This leads to the Potts
critical frontier
\begin{equation}
q^2+q(v+2w)-vw^2(3+w)=0,\quad B {\rm \>\>
lattice}. \label{mBPotts}
\end{equation}
Both expressions (\ref{mAPotts}) and (\ref{mBPotts}) are new.

\section{Percolation threshold}
 \medskip
We now specialize the above results to percolation.

\medskip
 It is well-known that bond percolation is realized by
taking the $q=1$ limit of the $q$-state Potts model with 2-site
interactions \cite{kf,wu2}. For  bond percolation on the martini lattice in Fig. 3,
we set $q=1$ and introduce bond occupation
probabilities $x_i=1-e^{-V_i}, \ y_i=1-e^{-W_i}$.  The percolation threshold 
 (\ref{3spin}) then assumes the form
 \begin{eqnarray}
&&x_1x_2(y_3+y_1y_2-y_1y_2y_3)+x_2x_3(y_1+y_2y_3-y_1y_2y_3)
          \nonumber\\
&& +\,x_3x_1(y_2+y_3y_1-y_1y_2y_3) 
-\,x_1x_2x_3(y_1y_2+y_2y_3+y_3y_1-2y_1y_2y_3) \nonumber \\
&=&1+(e^M-1)(1-x_1x_2-x_2x_3-x_3x_1+x_1x_2x_3) \, .\label{martinipergen}
\end{eqnarray}
For isotropic bond percolation $x_i=x,\,y_i=y$ and $M=0$, this reduces to
the threshold
\begin{equation}
3x^2y(1+y-y^2) -x^3y^2(3-2y)=1,\label{martiniper}
\end{equation}
which is a result obtained in \cite{ziff}. 

\medskip
For bond percolation on the martini
A and martini B lattices shown in Fig. 4, by setting $y_i=y$ and $x_i=x$ 
or $x_i=1$ (for $V_i=\infty$) we obtain  from
(\ref{martinipergen})  the thresholds
\begin{eqnarray}
 2xy(1+y-y^2)+x^2y(1-y)^2 &=&1, \quad\quad A {\rm \>\>
lattice},\nonumber \\
(1-y)^2(1+y) -xy(2-y) &=&0, \hskip 0.8cm B {\rm
\>\>lattice}.\label{abper}
\end{eqnarray}
 For uniform bond occupation probability $x=y=p$, (\ref{martiniper}) and
(\ref{abper})
 reduce to
  \begin{eqnarray}
  (2p^2-1)(p^4-3p^3+2p^2+1)&=&0, \ \quad {\rm martini\>\>lattice}, \nonumber \\
   p^5-4p^4+3p^3+2p^2&=&1,\ \quad A {\rm \>\>
  lattice},\nonumber \\
  (1-2p)(1+p-p^2)&=& 0, \quad \ B {\rm \>\>lattice},
  \end{eqnarray}
yielding the thresholds $p_c=1/\sqrt 2,\ 0.625457\cdots,$ and $1/2$
respectively.
 These numbers have been reported in \cite{scullard,ziff}.
Note that the thresholds (\ref{martiniper}) and (\ref{abper}) can
also be deduced from (\ref{mPotts}),  (\ref{mAPotts}),
and (\ref{mBPotts})  by setting $q=1$, $v=x/(1-x),\, w=y/(1-y)$.

\medskip
Consider next a correlated bond-site percolation process on the
honeycomb lattice with  edge   occupation probabilities
$x_1,x_2,x_3$ and alternate site occupation probabilities $s$ and
$1$. Now the site percolation is realized in the $q=1$ limit of
the $q$-state Potts model with multi-site interactions \cite{kw}.
Therefore, by setting
  $y_i=0$ and $s=1-e^{-M}$,  we obtain from
(\ref{martinipergen}) the critical frontier for this
site-bond percolation,
\begin{equation}
s(x_1x_2+x_2x_3+x_3x_1 -x_1x_2x_3)=1.\label{kondor}
\end{equation}
  The expression (\ref{kondor}), which generalizes an early result due to
Kondor \cite{kondor} for  $x_1=x_2=x_3$,
is the central result of \cite{scullard} derived from a
star-triangle consideration.
  Here, it is deduced as the result of an application of our
general formulation.

\medskip
As pointed out by Scullard \cite{scullard} and Ziff \cite{ziff},
the expression (\ref{kondor}) also gives the threshold for site
percolation on the martini lattice of Fig. 3, where $x_1,x_2,x_3$
are occupation probabilities of the three sites around a triangle and
$s$ is the occupation probability of the site at the center of the
$Y$.  For uniform occupation probability $x_1=x_2=x_3=s$,
(\ref{kondor}) yields the threshold $s_c =
0.764826\cdots$ for site percolation on the martini lattice \cite{scullard}.

\medskip
Setting $x_3=1$ in (\ref{kondor}) we obtain the threshold for site
percolation on the martini A lattice of Fig. 4(a) as
\begin{equation}
s(x_1+x_2)=1,\quad {\rm site\>\>percolation}-{\rm  A\>\>lattice}
\label{mAsite}
\end{equation}
where $x_1,x_2$ are occupation probabilities of the 3-coordinated
sites and $s$  the occupation probability of the 4-coordinated
sites.  For uniform occupation probability $x_1=x_2=s$,
(\ref{mAsite}) yields the threshold $s_c=1/\sqrt
2$ for site percolation on the A lattice. Likewise setting $x_2=x_3=1$ in
(\ref{kondor}), we obtain the threshold for site percolation on
the martini B lattice of Fig. 4(b),
\begin{equation}
s(1+x)=1,\quad {\rm site\>\>percolation}-{\rm  B\>\>lattice}
\label{mBsite}
\end{equation}
where $x=x_1$ and $s$ are, respectively, the occupation
probabilities of the 5-coordinated sites and 3-coordinated sites.
For uniform occupation probability $x=s$, (\ref{mBsite}) yields
the threshold $s_c=(\sqrt 5 -1)/2$ for site percolation on the B lattice.
These results have been reported in \cite{scullard,ziff}.

\section{Other lattices}

\medskip
As another example of our formulation, consider the Potts model on
the lattice in Fig. 5   with pure 2-site interactions $U,V,W\geq 0$.
  Writing $u=e^U-1,\ v=e^V-1,\ w=e^W-1$, we obtain
after a little algebra the Boltzmann factor (\ref{F}) with
\begin{eqnarray}
A&=& v^3 +3v^2(q+2w) +(3v+q)(q^2+3qw+3w^2+w^3) \nonumber \\
B&=&uA+v^2\Big[3w^2+w^3+(q+v)w\Big] \nonumber \\
C&=& u^2(u+3)A +3uv^2(u+1)(u+2)\Big[3w^2+w^3+(q+v)w\Big]\nonumber \\
&&  \hskip 2cm +(u+1)^3v^3(3w^2+w^3).
\end{eqnarray}
 The critical frontier
is again the self-dual trajectory $qA-C=0$.

\medskip
The resulting self-dual trajectory assumes a simpler form for the
percolation problem. For bond percolation we set $q=1$,
$u=z/(1-z), v=x/(1-x), w=y/(1-y)$ where $x,y,x$ are the respective
bond occupation probabilities shown in Fig. 5.  This yields the 
bond percolation critical threshold
\begin{eqnarray}
&&1-3z+z^3 -(1-z^2)\Big[3x^2y(1+y-y^2)(1+z) \nonumber
\\
&& \hskip 1cm +\,x^3y^2(3-2y)(1+2z)\Big]=0. \label{ex5}
\end{eqnarray}
Setting $z=0$ in (\ref{ex5}) it reduces to the bond percolation threshold
(\ref{martiniper}) of the martini lattice.
Setting $y=1$  (\ref{ex5})  gives the bond percolation threshold
 \begin{equation}
1-3z+z^3 -(1-z^2)\Big[3x^2(1+z)-x^3(1+2z)\Big]=0
 \label{ex51}
 \end{equation}
 for the dual of the martini lattice, which is the lattice in Fig. 5 with all
 small triangles shrunk into single points.

\medskip
For uniform bond percolation probabilities $x=y=z=p$, (\ref{ex5})
becomes
\begin{equation}
1-3p-2p^3+12p^5-5p^6-15p^7+15p^8-4p^9=0
\end{equation}
yielding the threshold $p_c=0.321808\cdots$.
Compared with the threshold $p_c=0.707106\cdots$ for the martini lattice,
it confirms the expectation that  percolation threshold
decreases
as  the lattice becomes more connected.

\section*{Summary and acknowledgment}
 \medskip
In summary, we have shown that the critical frontier of a host of
Potts models with 2- and multi-site interactions on lattices having the structure
 depicted in Fig. 1 can be explicitly determined. The resulting
critical frontier assumes the very simple form $qA-C=0$, where $A$ and $C$
are parameters defined in (\ref{F}). The corresponding threshold for bond and/or site
percolation are next deduced by setting $q=1$.
 Specializations of our formulation to the martini, the A, B, and
other lattices are presented. 

 \medskip
I would like to thank R. M. Ziff for sending me a copy of \cite{ziff}
prior to publication and for a useful conversation.  I am indebted to H. Y. Huang 
for  assistance in the preparation of the paper.

\newpage

\centerline{\bf Figure captions}
\bigskip
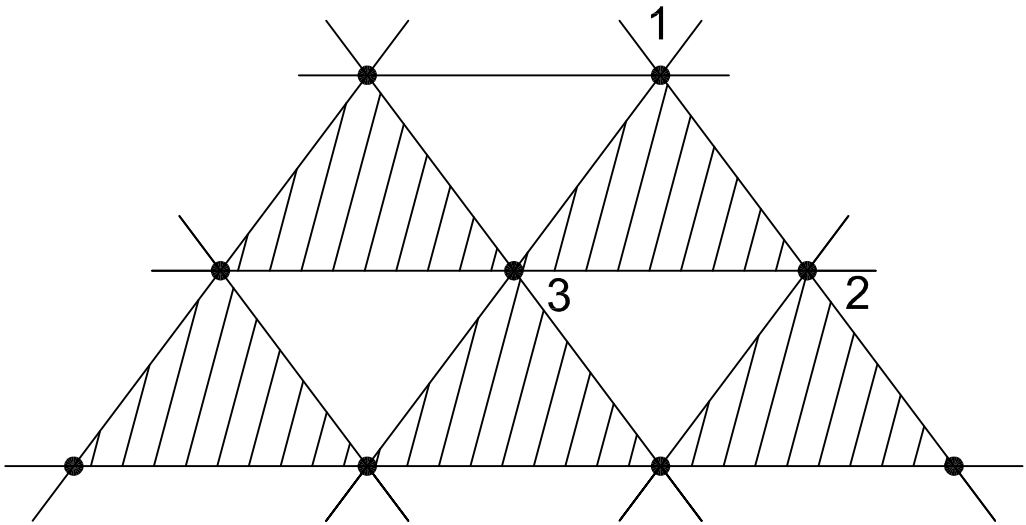
Fig. 1. The structure of a lattice possessing a duality relation.

\bigskip
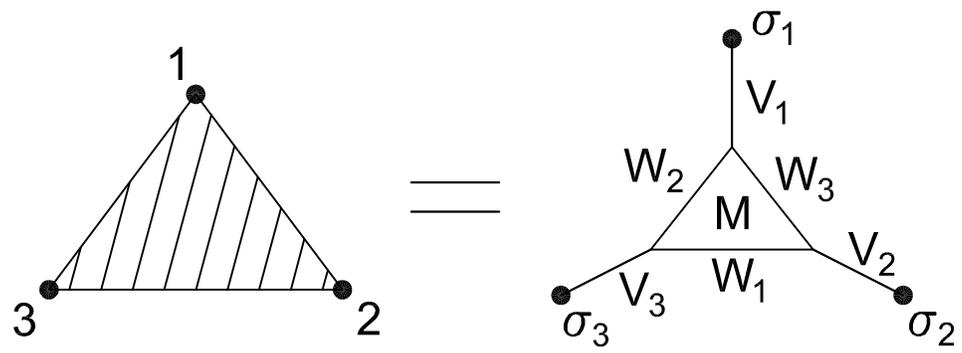
Fig. 2. The realization of Fig. 1 for the martini lattice.

\bigskip
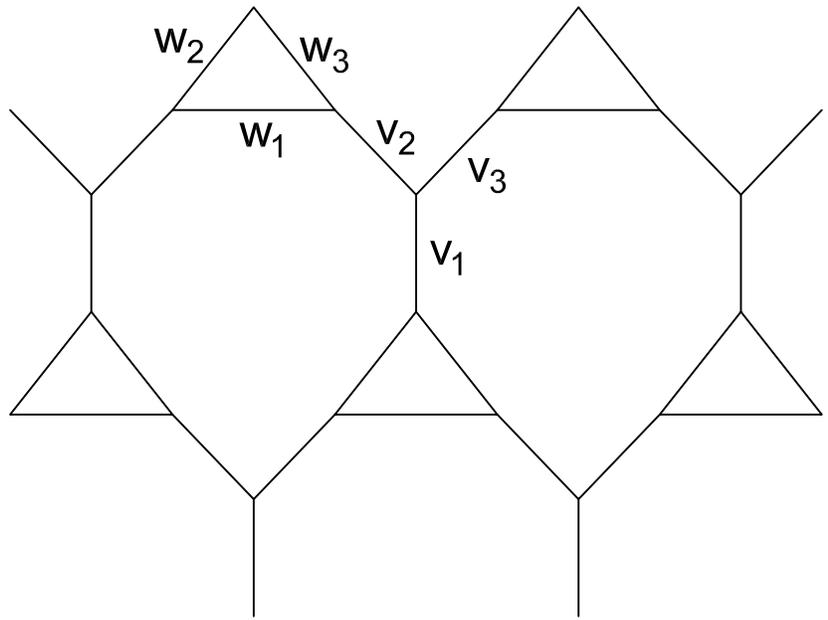
Fig. 3. The martini lattice.

\bigskip
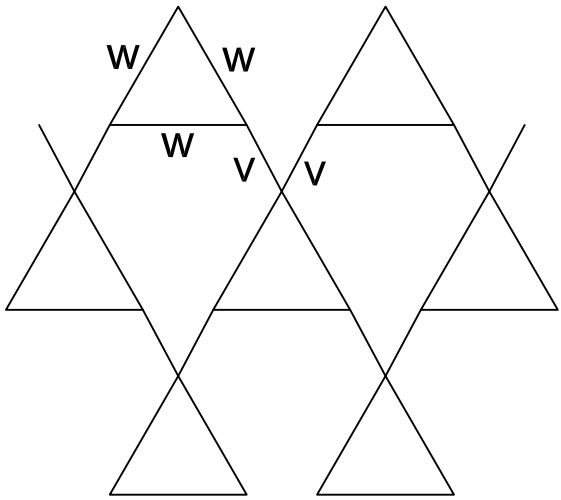
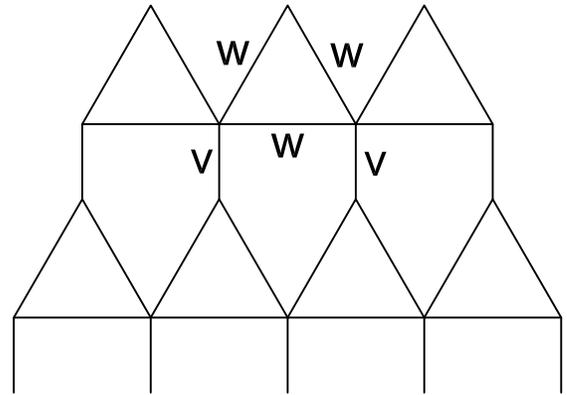
Fig. 4. (a) The martini-A lattice. (b) The martini-B lattice.

\bigskip
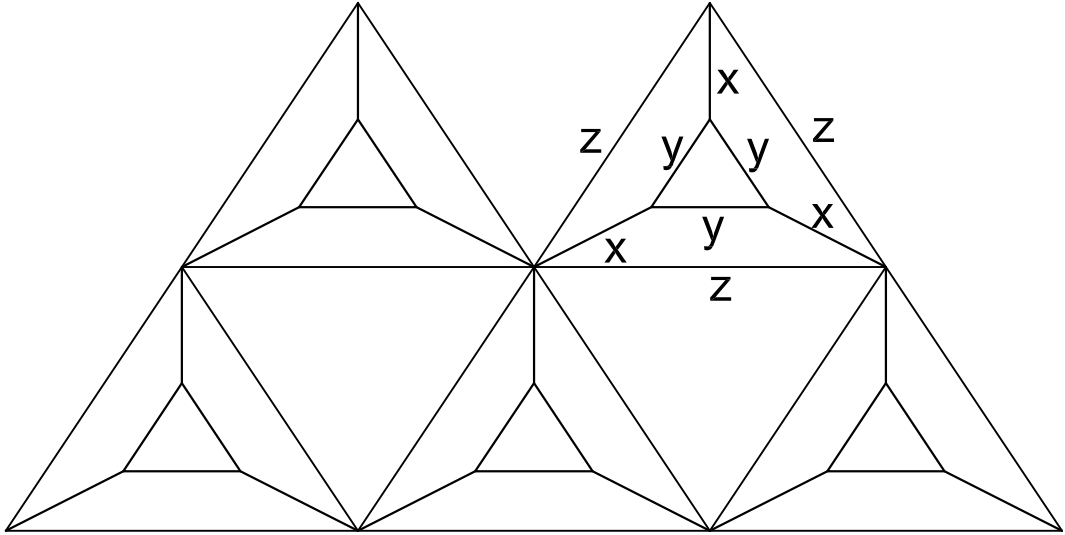
Fig. 5. A lattice with Potts interactions
$U,V,W$. Labels  shown are the corresponding
bond percolation probabilities\, $x=1-e^{-V},\ y=1-e^{-W},\ z=1-e^{-U}$.

 \end{document}